# Ge Quantum Dots Encapsulated by AlAs Grown by Molecular Beam Epitaxy on GaAs Without Extended Defects

Meng Qi, Chad A. Stephenson, Vladimir Protasenko, William A. O'Brien, Alexander Mintairov, Huili Grace Xing and Mark A. Wistey [a)]

*Electrical Engineering, University of Notre Dame, Notre Dame, Indiana, 46556, USA*

**Abstract**

We demonstrate nearly-spherical, strain-free, self-assembled Ge quantum dots (QDs) fully encapsulated by AlAs, grown on (100) GaAs by molecular beam epitaxy (MBE). The QDs were formed without a wetting layer using a high temperature, in-situ anneal. Subsequent AlAs overgrowth was free from threading dislocations and anti-phase domains. The straddling band alignment for Ge in AlAs promises strong and tunable confinement for both electrons and holes. The reflection high-energy electron diffraction (RHEED) pattern changed from 2×3 to 2×5 with anneal, which can be explained by surface reconstructions based on the electron-counting model.

**Keywords:** Ge quantum dots, molecular beam epitaxy, surface reconstruction, carrier confinement

Germanium quantum dots (QDs) have interesting properties such as a large exciton binding energy, long radiative lifetime, and strong size dependence of radiative lifetime,[1] which could be utilized for optical devices. In particular, as an indirect bandgap material, Ge can offer long minority carrier lifetimes, ranging from μs to ms.[2,3] However, carrier confinement in Ge nanostructures is difficult without lattice matched barrier materials. Ge

---

[a)]Electronic mail: mwistey@nd.edu;   Tel: (574) 631-1639;   Fax: (574) 631-4393

QDs reported to date have nearly all been grown on Si, with QD formation occurring by a Stranski-Krastanov (SK) growth mode driven by the 4% strain between Ge and Si.[4,5,6,7] But Si barriers provide relatively weak confinement for holes and none for electrons. Furthermore, SK growth leaves a wetting layer that prevents subsequent III-V layers from being registered to the underlying substrate. This can lead to anti-phase domains (APDs) due to the growth of polar III-V's on nonpolar Ge.[8,9,10,11,12] Ge nano-crystals in a dielectric matrix have been well studied and show different PL emission at visible light range.[13,14,15,16] Such emission is attributed to excitonic emission in a quantum-confined system.[Error! Bookmark not defined.,17,18] But this technique shows poor interfaces between the dielectric and the Ge nano-crystals, which can influence the optical emission properties.[19,20] Emission in the near infrared has also been demonstrated,[21,22] and size-dependent emission has been theoretically calculated.[23]

In this work, self-assembled Ge QDs were formed directly on (100) AlAs by thin (0.55 or 1.1 monolayer (ML)) Ge growth followed by in-situ high temperature annealing. Ge and AlAs are nearly lattice-matched, so QD ripening is not primarily driven by strain. Due to the Type-I (straddling) band alignment of the AlAs/Ge heterojunction and the lack of a wetting layer, this technique offers a route to defect-free growth of an unusual III-V/IV/III-V porous superlattice with strong confinement of both electrons and holes. The nearly spherical shape of the fully encapsulated Ge QDs in all three dimensions increases the effects of quantum confinement. This novel structure has potential applications for up-conversion solar cells, light emitting devices, single electron tunneling devices, and other optoelectronics.

The growths were carried out in several interconnected MBE chambers including a Veeco Gen 930 for III-V semiconductors and an Intevac Mod Gen II for Group IV semiconductors. The growth temperatures were calibrated by a kSA BandiT system, which simultaneously measures temperature by pyrometry, band edge optical absorption, and blackbody radiation curves. As$_2$ was provided by a Veeco Mark V valved cracker. The growth rates were 1 μm/hour for GaAs and AlAs and 600 nm/hour for Ge. In this growth, semi-insulating (100) GaAs substrates were heated to 630 °C for 15 min to fully desorb surface oxides. Then 200 nm of non-intentionally-doped GaAs and 200 nm AlAs were grown at 610 °C with a V/III beam equivalent pressure ratio of 20. The sample was then cooled to room temperature, with As flux maintained until below 400 °C to prevent decomposition. The sample was transferred under UHV to the Group IV chamber for solid source Ge growth. 0.55 monolayer (ML) and 1.1 ML of Ge were grown at 410 °C for different samples. The sample was then transferred back to the III-V chamber for the annealing at 690 °C. A beam equivalent pressure of As$_2$ = 6.7×10$^{-6}$ Torr was used to prevent decomposition of the AlAs above 400 °C during the anneal. After the annealing, several samples were overgrown with cap layers of AlAs and GaAs. To minimize oxidation, each sample was unloaded quickly and immediately placed in an Ar purged glovebag. Uncapped samples were then loaded under inert gas into an N$_2$-purged atomic force microscope (AFM) to explore the surface morphology, as shown in Fig. 1. In addition, transmission electron microscopy (TEM) specimens were prepared by focused ion beam (FIB). Cross-section high resolution TEM (X-HRTEM) and scanning TEM (STEM) results are shown in Fig. 2.

AFM of uncapped samples confirmed the formation of Ge QDs, as shown in Fig. 1. A control sample had 0.55 ML initial Ge on AlAs without annealing, which showed a smooth surface without QDs formation. Deposition and annealing of 0.55 ML Ge formed QDs with a density of 6×10$^9$ cm$^{-2}$ after 10 min at 690 °C, while 1.1 ML formed larger QDs with density of 2.8×10$^{10}$ cm$^{-2}$. For the 1.1 ML initial Ge layer, a longer anneal of 14 minutes at the same temperature, increased the size and reduced the density of QDs to 5.6×10$^9$ cm$^{-2}$, consistent with Ostwald ripening,[24,25] which has demonstrated similar island formation on silicon.[26] This indicates the growth mechanism is surface energy driven.

The absence of a wetting layer provides ample exposed III-V surface for subsequent growth of additional AlAs without antiphase domains (APDs). The RHEED during the overgrowth to encapsulate Ge QDs showed the same 2×4 and faint 2×4 pattern as normal GaAs and AlAs growth respectively. HRTEM (Fig. 2) shows that the Ge QDs are fully encapsulated in AlAs, with defect-free AlAs and GaAs subsequent layers grown on top. No threading dislocations or APDs were visible in any observation. The inset Z-contrast image shows a single Ge QD surrounded by AlAs. The lattice structure can be observed, along with a lateral size of 7 nm and vertical height of 4.5 nm. The height from HRTEM is consistent with the height from AFM, while AFM shows a lateral diameter of 15-25 nm, which maybe limited by the geometry of the AFM tip.

Raman spectroscopy was used to characterize Ge content in the QDs. In Fig. 3, the inset shows the structure of the sample for measurement. A 1 μm $Al_{0.4}Ga_{0.6}As$ buffer followed by 1 μm AlAs was grown on GaAs in order to absorb pump laser and thus prevent a Raman peak from the GaAs substrate, since the Raman peak of bulk GaAs at 291.5 $cm^{-1}$ is close enough to that of bulk Ge at 300.5 $cm^{-1}$ that they could overlap. A 5 nm $Al_{0.4}Ga_{0.6}As$ cap layer on top of four periods of 1.1 ML Ge QDs/15 nm AlAs was used to prevent the rapid oxidation of AlAs. Raman spectroscopy showed a peak at 299 $cm^{-1}$, which we attribute to the LO phonon energy of Ge, 300.5 $cm^{-1}$, shifted about 1.5 $cm^{-1}$. The downshift of the Ge Raman peak was likely caused by sample heating from the Raman pump laser, based on power-dependent measurements (not shown).[27] Also, the line shape of the Ge QDs was observed to be asymmetric. Similar asymmetric line shape has been reported for InAs/AlAs QDs and other arsenide material systems.[28] A similar asymmetric phonon line shape of Ge QDs is characteristic of Raman spectra of nanocrystalline structures, which can be described by a model of phonon confinement in nanoclusters of inhomogeneous size.[29,30] The clear Ge peak in Raman spectroscopy suggests that the Ge QDs are at least relatively pure.

The RHEED patterns before and after annealing at the same temperature as well as during annealing at 690 °C are shown in Fig. 4. The conventional 2×4 RHEED pattern for AlAs was not observed during the annealing. Before annealing, a streaky 2×3 RHEED reconstruction pattern was observed for the 0.55 ML Ge/AlAs surface at 610 °C, as shown in upper figure in Fig. 4. During the annealing, the 2×3 pattern becomes more diffusive and the streaky lines become less continuous, which are consistent with QDs formation. After annealing, a unique 1×5 or weak 2×5 was observed. In order to correlate the RHEED pattern to the QDs formation mechanism, a surface reconstruction model is proposed in Fig. 5, based on the electron-counting model.[31] At relative low temperature before annealing, the initial Ge layer is uniformly distributed, confirmed by the streaky 2×3 RHEED pattern and AFM (Fig. 1(a)). Given the fact that Ge is an n-type dopant in bulk AlAs, it is more preferable for Ge to replace Al than As. With a single Ge-Al substitution in the 2×3 unit cell shown in Fig. 5 (a), the cell fulfills the electron-counting rule and becomes a stable surface reconstruction. Also, the 2×3 reconstruction arises from the one missing As-As dimer for every two dimers. With more Ge-Al and Ge-As substitutions still makes possible stable surface reconstructions based on electron-counting rule, as shown in Fig. 5(c). During the annealing at around 700 °C, the Ge atoms become more mobile due to the higher kinetic energy, and we believe the surface reconstruction favors Ge-Ge bonds over Ge-As bonds. Fig. 5 (b) shows a 2×3 unit cell with a Ge-Ge dimer on top still fulfilling electron-counting model. Possible stable 2×3 surface reconstruction models with different Ge concentration, as shown in Fig. 5(a)-(c), explained how 2×3 pattern was observed for sample with different initial Ge thickness of 0.55 ML and 1.1 ML. The anneal changes the Ge distribution, and most of the surface has much less Ge coverage due to the QD formation. With one Al atom replaced by Ge, the 2×5 becomes a stable structure, as shown in Fig. 5 (d). This is consistent with the less Ge coverage in most of the area other than QDs area. The high temperature annealing also makes the long term ordering available thus forming the 2×5 surface reconstruction as observed in RHEED. Further experiments will be performed to test this theory by altering the surface and formation energies of QDs using various terminating species.

We believe the self-assembly of Ge QDs on AlAs is driven by a high surface energy caused by Ge atoms avoiding the As bonds at high temperature.[31] At lower temperatures such as during Ge deposition, Ge atoms have low surface mobility, and As is similarly stable in its bonds with Al. This prevents atomic rearrangement but leads to an energetically unfavorable surface due to violation of electron counting.[32] During annealing, however, Ge reduces its surface energy by forming Ge dots with a high aspect ratio, and removing most of the Ge from the surface. The lack of a wetting layer is due to the strain-free relaxation mechanism, since there is an energetic penalty for the first monolayer due to broken bonds, but no additional penalty for additional monolayers. The observation of Ostwald ripening as shown in Fig. 1 suggests that QDs are driven toward a lower surface-to-volume ratio with larger, more stable QDs. We believe this indicates atoms on the QD surfaces are especially unfavorable due to dangling bonds, which are only partly relieved by reconstruction or bonded excess As. Based on the Al-As-Ge alloy phase diagram, AlAs has a solubility limit less than 2% in Ge at around 700 °C, and a eutectic temperature at 735 °C.[33,34,35] Although there is a Al-Ge eutectic point at 420-425 °C,[36] no change is visible in RHEED until near the annealing temperature of 690 °C. It could due to the fact that Ge and Al have insufficient surface mobility at low temperature to gather into droplets, or the presence of arsenic may suppress the Ge-Al eutectic formation. At annealing temperature, we believe the presence of the surface reduces the eutectic temperature of Ge-AlAs, and the alloy becomes droplets as Al is dissolved from the AlAs layer. When the substrate is cooled under $As_2$, the alloy may segregate into AlAs and Ge, but it retains the circular shape and high aspect ratio of droplets. Better recrystallization conditions, such as cooling slowly through the freezing point under As rich conditions, may increase the crystal quality by favoring the growth of AlAs as AlAs segregates from the melt.

In summary, we have demonstrated self-assembled Ge QDs grown on AlAs. The QDs were formed without a wetting layer, using a high temperature annealing step, and the lack of a wetting layer allowed subsequent growth of additional AlAs without APDs to fully encapsulate the Ge QDs. Raman spectroscopy confirmed the high-purity Ge based on a peak at 299 cm$^{-1}$. The downshift of 1.5 cm$^{-1}$ from the bulk Ge Raman peak for the Ge QD phonon peak is attributed to thermal heating during measurement. And the asymmetric line shape could be caused by the inhomogeneous sizes of the QDs as well as the phonon confinement. Ostwald ripening was observed, which confirmed that self-assembly growth mechanism of Ge QDs on AlAs originates from the surface free energy and surface tension of Ge on AlAs at the annealing condition. RHEED patterns of 2×3 and 2×5, uncommon in AlAs, were observed before and after the Ge QDs formation, respectively. A surface reconstruction model based on electron-counting rule was proposed and used to explain the RHEED. This technique can offer an APD-free combination of Group IV and III-V materials, as well as direct and indirect bandgaps (if AlGaAs is used), for new device applications.

**Acknowledgement**

The authors thank Dr. Xinyu Liu, James Kapaldo, Guowang Li and Pei Zhao at University of Notre Dame for helpful discussions.

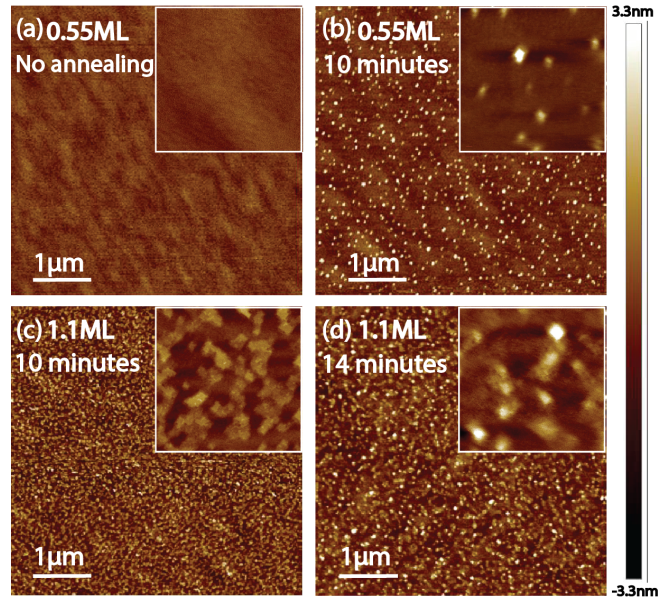

Fig. 1: AFM under inert gas of uncapped Ge QDs on AlAs, with 690 °C anneal. (a) 0.55 ML of Ge without annealing, (b) 0.55 ML Ge after 10 min anneal, (c) 1.1 ML Ge with 10 min anneal, (d) 1.1 ML Ge with 14 min anneal. Insets are higher magnification 500 nm by 500 nm scan.

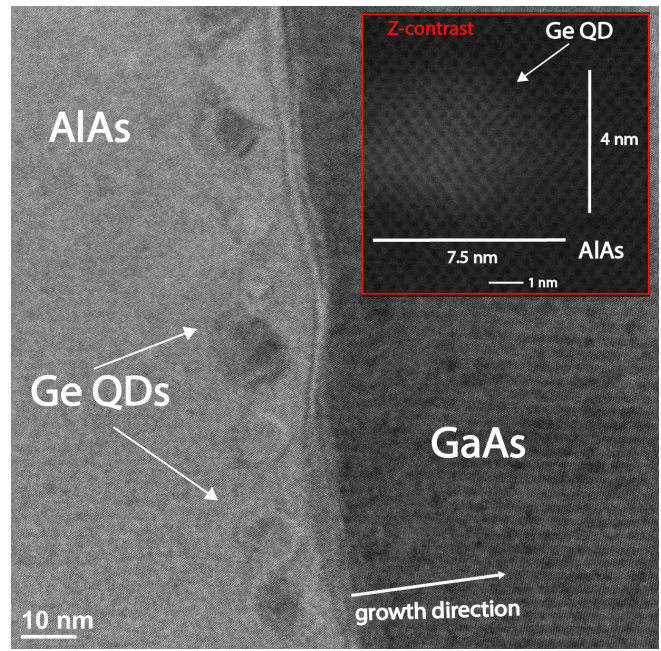

Fig. 2: X-HRTEM shows QDs formed by annealing of 1.1 ML Ge on AlAs, followed by subsequent AlAs and GaAs overgrowth. The inset shows a Z-contrast STEM lattice image. Lattice of Ge QDs with higher contrast can be observed with AlAs lattice around. The dimensions of a single Ge QD are measured in the inset figure. Note lack of a wetting layer in STEM mode.

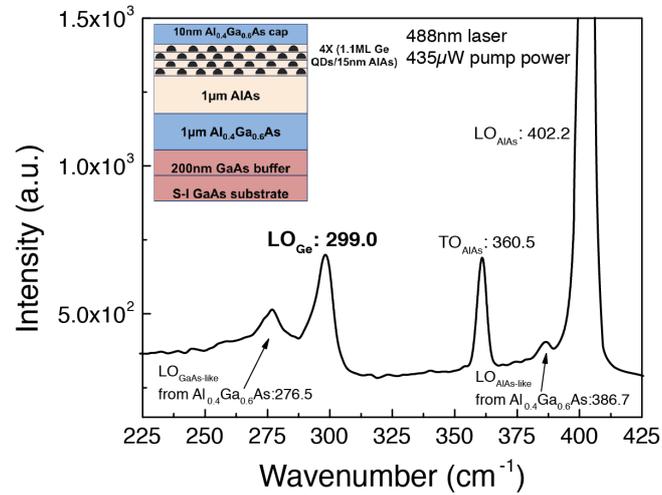

Fig. 3. Raman spectroscopy of Ge QDs. The LO phonon peak of Ge is observed at 299 cm$^{-1}$. Other peaks[37] from the $Al_{0.4}Ga_{0.6}As$ and AlAs buffer are also labeled. The sample structure is shown schematically in the inset.

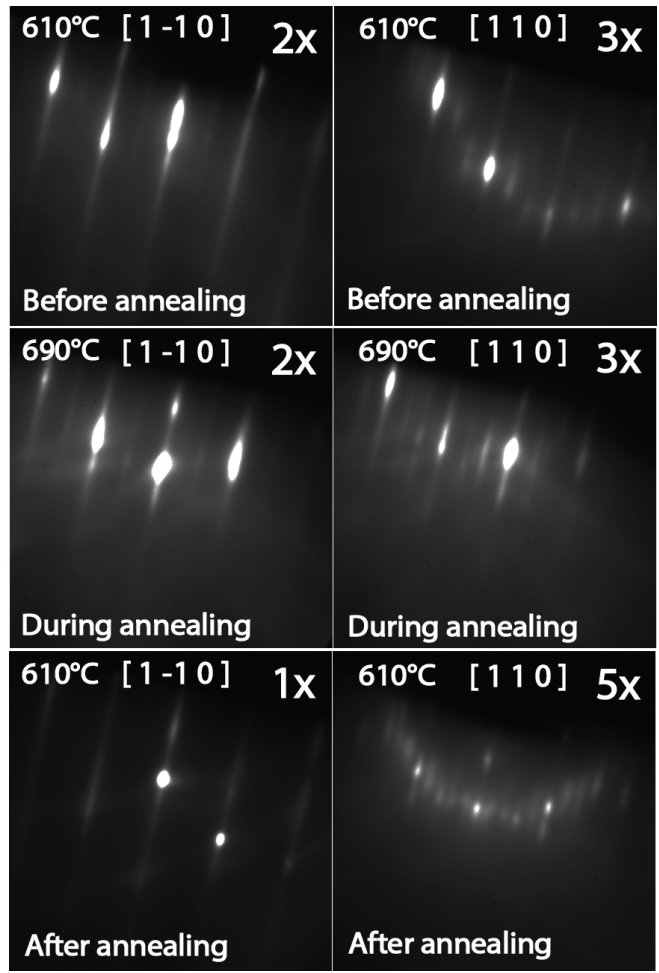

Fig. 4. RHEED patterns observed at substrate temperatures of 610 °C before annealing (top row), 690 °C during annealing (middle row) and 610 °C after annealing (bottom row). 0.55ML of Ge was deposited on AlAs before annealing. A 2×3 pattern was observed before and during annealing since there is one weak line between the main lines at [1 -1 0] and two weak lines between mainlines at [1 1 0]. Some of the samples show weak 2×5 after the annealing at 610°C instead of 1×5 under the same condition. All the samples were grown on ( 0 0 1) GaAs substrates.

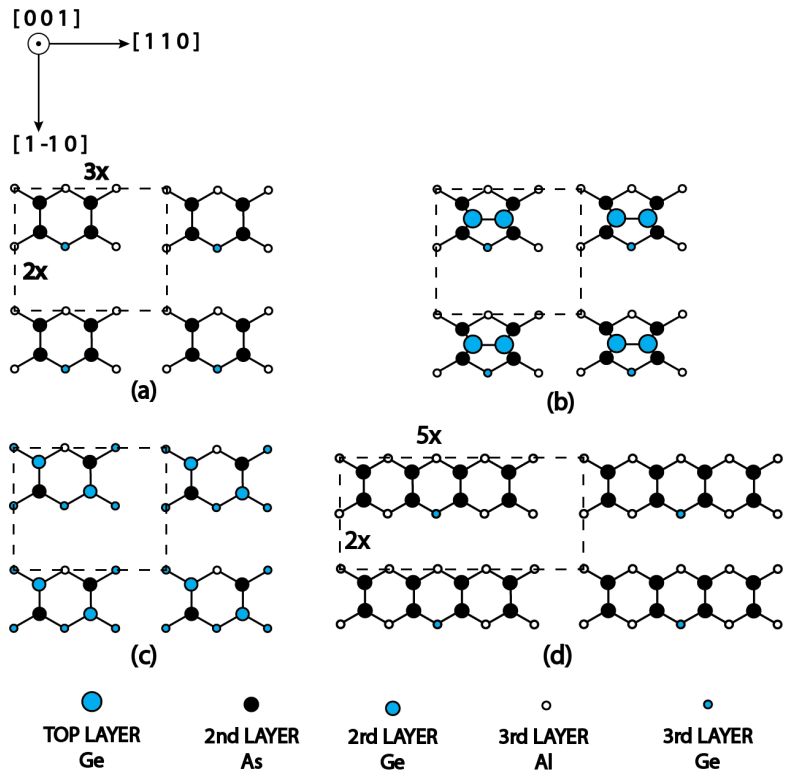

Fig. 5. Unit cells of surface reconstruction model based on electron counting model for 2×3 (a)-(c) and 2×5 (d) RHEED patterns. (a) and (b) show possible surface reconstructions before annealing in areas with <1 ML Ge, one Al atom is replaced by Ge atom in each unit to fulfill electron-counting rules; (b) shows unit cell with a Ge-Ge dimer on top which gives a higher Ge concentration; (c) shows possible surface reconstruction before annealing with higher Ge content before annealing with >1ML Ge, five Al and two As atoms are replaced by Ge atoms in each unit cell; (d) shows the surface reconstruction of non-QD area after annealing with a much lower Ge concentration. Only one example of possible unit cell structures is shown for each situation.

International, Materials Park, OH, USA, 2006-2013

[35] J.H. Bryden, *Acta Crystallogr.* 15, 167 (1962)

[36] Facility for the Analysis of Chemical Thermodynamics (F*A*C*T), Centre for Research in Computational Thermochemistry, École Polytechnique de Montréal. URL: http://www.crct.polymtl.ca/fact/phase_diagram.php?file=Al-Ge.jpg&dir=SGTE

[37] G. S. Solomon, D. Kirillov, H. C. Chui, and J. S. Harris, *J. Vac. Sci. Technol. B* 12, 1078 (1994)